\documentclass[aps,prl,reprint,showpacs,twocolumn,superscriptaddress, footinbib,preprintnumbers,floatfix]{revtex4-1}
\usepackage{amsmath,amssymb}
\usepackage{times}
\usepackage{graphicx}
\usepackage{natbib}
\usepackage{hyperref}
\usepackage{bm}
\usepackage{slashed}
\usepackage{xcolor}
\usepackage{verbatim}

\newcommand{\bw}{\begin{widetext}}
\newcommand{\ew}{\end{widetext}}
\newcommand{\be}{\begin{equation}}
\newcommand{\ee}{\end{equation}}
\newcommand{\bestar}{\begin{equation*}}
\newcommand{\eestar}{\end{equation*}}

\newcommand{\bi}{\begin{itemize}}
\newcommand{\ei}{\end{itemize}}
\newcommand{\bea}{\begin{eqnarray}}
\newcommand{\eea}{\end{eqnarray}}

\newcommand{\hbo}{\hbox to 1 true cm {\hfill } }

\newcommand{\vcb}[1]{\mbox{\bf #1}}

\newcommand{\ud}{\mathrm{d}}

\newcommand{\m}{{\scriptscriptstyle -}} 
\newcommand{\p}{{\scriptscriptstyle +}}
\newcommand{\LCperp}{{\scriptscriptstyle \perp}}

\newcommand{\sfrac}[2]{{\textstyle \frac{#1}{#2}}}
\newcommand{\phii}{\phi_{\mathrm{i}}}
\newcommand{\phif}{\phi_{\mathrm{f}}}
\newcommand{\dotp}[2]{#1 \! \cdot \! #2}
\newcommand{\er}{\epsilon_{\mathrm{rad}}}
\begin{document}
\title{Detecting radiation reaction at moderate laser intensities}
 
\author{Thomas Heinzl}
\email{theinzl@plymouth.ac.uk}
\affiliation{School of Computing and Mathematics, Plymouth University, Plymouth PL4 8AA, UK}

\author{Chris Harvey}
\affiliation{Centre for Plasma Physics, Queen's University Belfast, BT7 1NN, UK}

\author{Anton Ilderton}
\affiliation{Department of Applied Physics, Chalmers University of Technology, SE-41296 Gothenburg, Sweden}

\author{Mattias Marklund}
\affiliation{Department of Applied Physics, Chalmers University of Technology, SE-41296 Gothenburg, Sweden}
\affiliation{Department of Physics, Ume\aa\ University, SE-901 87 Ume\aa, Sweden}

\author{Stepan~S.\ Bulanov}
\affiliation{University of California, Berkeley, California 94720, USA}

\author{Sergey Rykovanov}
\affiliation{Lawrence Berkeley National Laboratory, Berkeley, California 94720, USA}

\author{Carl B. Schroeder}
\affiliation{Lawrence Berkeley National Laboratory, Berkeley, California 94720, USA}

\author{Eric Esarey}
\affiliation{Lawrence Berkeley National Laboratory, Berkeley, California 94720, USA}

\author{Wim P. Leemans}
\affiliation{Lawrence Berkeley National Laboratory, Berkeley, California 94720, USA}

\begin{abstract}

We propose a new method of detecting radiation reaction effects in the motion of particles subjected to laser pulses of moderate intensity and long duration. The effect becomes sizeable for particles that gain almost no energy through the interaction with the laser pulse.

\end{abstract}


\maketitle

\paragraph{Introduction:--}
In a conceptually simple experiment~\cite{Meyerhofer:1996} it was shown that electron motion in a sufficiently intense laser becomes relativistic. In that experiment, a laser pulse was used to ionise a target gas, liberating electrons. After the electrons left the pulse, their energies and ejection angle were measured. Different values for these variables are predicted by relativistic and non-relativistic equations of motion; the experiment supported the relativistic prediction. (The phrase `mass shift' in~\cite{Meyerhofer:1996} refers to the `relativistic mass' $m\gamma$; the experiment was not concerned with, and did not observe, the intensity dependent mass shift, for which see~\cite{Harvey:2012ie}.)

In this paper we propose a similar experiment to measure classical radiation reaction (RR). The problem of RR on the dynamics of charged particles in electromagnetic (EM) fields is long standing, and has attracted a lot of attention for more than a century. It is relevant for charged particle acceleration in terrestrial laboratories and in ultra-high energy cosmic rays. The interaction of charged particles with laser radiation provides special conditions for studying not only the interaction itself, but also RR effects. Present day PW-class laser facilities, such as BELLA \cite{BELLA}, are at the threshold of the interaction regime dominated by RR effects, which are potentially able to completely change the nature of charged particle interactions with EM fields~\cite{DiPiazza:2011tq,control,Gonoskov:2013aoa}.   

The idea of this paper is simple: the same experiment as in~\cite{Meyerhofer:1996} is performed, and the properties of the emitted electrons measured. These are then used to test the predictions of the classical equations of motion with and without RR. There is no need to measure the properties of the produced radiation. This is good news in view of the recent finding (for a different interaction set-up) that RR effects are almost invisible in the radiation spectrum while they are more than obvious in the electron distribution~\cite{Thomas:2012}. This difference in size is consistent with the fact that RR effects are suppressed in the photon spectrum (by a factor of the classical RR parameter, see below) relative to those in the electron spectrum~\cite{Ilderton:2013tb}.
\paragraph{Review:--}
Let the laser propagate along the $z$-axis. The polar and azimuthal electron ejection angles, relative to this axis, are $\theta\in\{0,\pi\}$, $\varphi\in\{0,2\pi\}$ respectively. They are determined by the following electron velocity ratios at large times, i.e.\ after the pulse has passed ($\perp=\{x,y\}$),
\be \label{TAN}
 \tan \varphi = \frac{u^y}{u^x}    \; , \qquad  \tan \theta = \frac{|u^\LCperp|}{u^z} \; .
\ee
In the original experiment \cite{Meyerhofer:1996}, the polar angle measurement was accompanied by a determination of the electron energy, i.e.\ its gamma factor, $\gamma = E_p/m$. The laser had a pulse duration of $1$ ps and a peak intensity of approximately 10$^{18}$ W/cm$^2$.  The experimental results were compared against the theoretical analysis of~\cite{Meyerhofer:1996} which assumed the laser to be a plane wave. For propagation along the $z$-axis, the plane wave depends on the invariant phase $\phi := \dotp{k}{x}$ where the four-momentum $k^\mu = \omega(1,0,0,1)\equiv \omega n^\mu$ is lightlike and $\omega$ is a typical frequency. In a plane wave, a charge's velocity component $\dotp{n}{u}=\gamma-u^z\equiv u^\m$ is conserved~\cite{Troha:1999pg,Harvey:2011dp}, as is the transverse canonical momentum; this allows the remaining $u^\p$ component to be determined by the mass-shell condition. Let the pulse extend over $0\leq \phi\leq \phif$ and let an electron `appear' in the pulse at phase $\phii$, with velocity $u_\mathrm{i}$, following ionisation. The particle's subsequent velocity takes on the compact form
\be \label{LORENTZ.SOLN}
  u^\mu = u_\mathrm{i}^\mu - a^\mu + (\dotp{u_\mathrm{i}}{a} - a^2/2) \, \frac{n^\mu}{\dotp{n}{u}} \;,
\ee 
in which dimensionless $a^\mu$ is the phase integral of the (tranverse) electric field $E^\mu \equiv (0, \vcb{E}_\LCperp, 0)$, from the initial time and in relativistic units,
\be \label{A.INTEGRAL}
   a^\mu (\phii; \phi) =  \int_{\phii}^\phi \ud \varphi \, \frac{eE^\mu(\phi)}{m\omega} \;.
\ee 
We refer to this as the potential~\cite{Foot1}. We have deliberately made explicit the dependence on the phase value $\phii$ at ionisation. For the rest of the paper we will also assume, as in~\cite{Meyerhofer:1996}, that the electron is at rest immediately post-ionisation, $u_\mathrm{i}^\mu=(1,0,0,0)$, which is a natural approximation for ionisation by a linearly polarised EM wave \cite{Popov}.

With the above assumptions, one finds that the polar emission angle $\theta(\phii;\phif)$, evaluated at the final phase $\phi = \phif$ marking the end of the pulse, obeys
\be \label{TAN.FTM1.5}
	\tan\theta \, (\phii;\phif) = \frac{2}{|a_\LCperp(\phii;\phif)|}\; ,
\ee
and that this is correlated with the final gamma factor by
\be \label{TAN.FTM1}
  \tan^2\theta \, (\phii;\phif) = \frac{2}{\gamma(\phii;\phif) - 1} \;.
\ee
This parametric relation was tested and confirmed in~\cite{Meyerhofer:1996}, for a variety of targets giving different ionisation times $\phi_\mathrm{i}$.

\begin{figure}[t!]
\includegraphics[width=0.8\columnwidth]{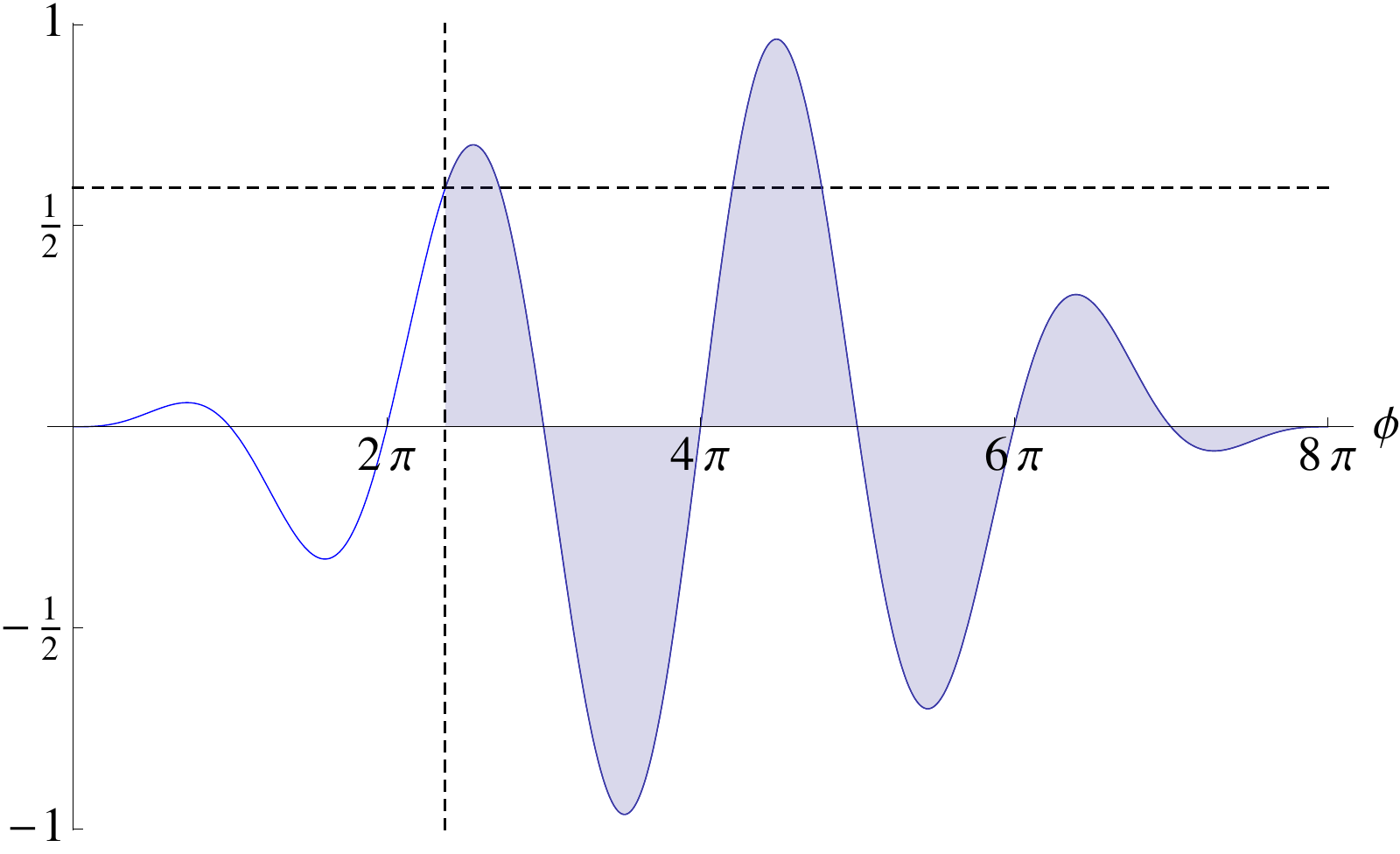}
\caption{\label{FIG:PULSE} Sketch of the field of a laser pulse. The horizontal dashed line represents the ionisation threshold. Only the shaded region contributes to the pulse integrals (\ref{A.INTEGRAL}) determining the electron ejection angle.}
\end{figure}

From (\ref{TAN.FTM1}), the ejection angle measures the energy transfer to the electron in a plane wave. That this is non-zero does not contradict the Lawson-Woodward theorem~\cite{woodward:1946,woodward:1948,palmer:1987}. The loophole is that the electrons do not see the whole pulse; they are bound in atoms until the pulse's amplitude exceeds the ionisation threshold, at which point, $\phi_\mathrm{i}$, they are injected into the pulse, see Fig.~{\ref{FIG:PULSE}}. The energy transfer predicted in (\ref{TAN.FTM1}) and confirmed in \cite{Meyerhofer:1996} is therefore an example of ionisation induced sub-cycle acceleration~\cite{Pantell:1978,Plettner:2005jd}. Had the electron seen the whole pulse, its net energy-momentum gain would have been zero because $a(0;\phif)=0$, assuming the background has no DC-component~\cite{Dinu:2012tj}.

\paragraph{Radiation Reaction:--}
RR terms in the Lorentz-Abraham-Dirac equation~\cite{Lorentz:1906,Abraham:1905,Dirac:1938} appear multiplied by the purely classical time parameter (temporarily reinstating $c$)
\be
  \tau_0 := \frac{2}{3} \frac{r_e}{c} = \frac{e^2}{6\pi mc^3} = \frac{2}{3} \alpha
  \frac{\lambdabar_e}{c} \simeq 3 \; \mbox{fm}/c \; ,
\ee
$r_e$ denoting the classical electron radius, $\alpha = e^2/4\pi \hbar c \simeq 1/137$ the fine structure constant and $\lambdabar_e$ the Compton wavelength of the electron. A dimensionless parameter $\er$ characterising RR may be obtained by taking the ratio of $\tau_0$ to the typical time scale of the laser, $1/\omega$: 
\be \label{r}
  \er := \omega \tau_0 = \frac{2}{3}\frac{r_e}{\lambdabar} = \frac{2}{3}
  \alpha\frac{\hbar\omega}{mc^2} ,
\ee
with $\lambdabar$ the (reduced) laser wavelength. A precursor of this parameter was already introduced by Lorentz \cite{Lorentz:1906}, see the useful overview article \cite{McDonald:2000} and Koga et al.\ emphasised its importance in a discussion of RR corrections to nonlinear Thomson scattering \cite{Koga:2005}. When $\er$ approaches unity one reaches a regime where the RR force is of the same magnitude as the Lorentz force, but as $\hbar \omega \simeq 200 \, mc^2$ in this case, one has simultaneously entered the quantum regime~\cite{BulanovNIMA:2011}.
 
In this paper we will treat RR as a correction to the Lorentz force effects, i.e.\ we will work to first order in $\er$. To this (and only this) order the Lorentz-Abraham-Dirac and Landau-Lifshitz equations~\cite{Landau:1987} are identical. (See~\cite{Bulanov:2011} for a recent comparison.) We can therefore appeal to the known analytic solution of the Landau-Lifshitz equation in a plane wave~\cite{diPiazza:2008}, and then truncate to order $\er$. The $\mathcal{O}(\er)$ expressions are not illuminating, so for simplicity we recall here the exact solution, which may be written akin to the Lorentz solution~(\ref{LORENTZ.SOLN}). Following~\cite{diPiazza:2008}, we introduce
\be \label{H}
	h(\phii; \phi) = 1 - \er \int\limits_{\phii}^{\phi} \ud \varphi \, a^{\prime 2} \;, 
\ee
which parameterises the main dynamical effect of RR on a particle in a plane wave, that being that $u^\m$ ceases to be conserved~\cite{Troha:1999pg,Harvey:2011dp}. One has instead $u^\m (\phii;\phi) = u_{\mathrm{i}}^\m / h(\phii; \phi)$, which is monotonically decreasing. For a particle initially at rest, and abbreviating $a' \equiv eE/m\omega$, the solution of the Landau-Lifshitz equation assumes the compact form
\be \label{LL.SOLN}
 h u^\mu = u_{\mathrm{i}}^\mu - \mathcal{A}^\mu + \left[ -\sfrac{1}{2} \mathcal{A}^2 + \sfrac{1}{2} (h^2 - 1) \right] \frac{n^\mu}{\dotp{n}{u}} \; ,
\ee
with the modified potential
\be \label{A.RR}
  \mathcal{A}^\mu := \int\limits_{\phii}^\phi \ud \varphi \, (h a^{\prime \mu} + \er a^{\prime \prime \mu}) \; .
\ee
$\mathcal{A} \to a$ in the absence of RR, i.e.\ when $\er \to 0$.  The essential point is simply that the predictions of (\ref{H})-(\ref{A.RR}) are quantitatively different from those of (\ref{LORENTZ.SOLN}), so that an experiment like that in \cite{Meyerhofer:1996} can in principle be used to detect RR effects. 

It is here convenient to factorise the electric field into amplitude, shape and polarisation. We therefore define $a_0 = eE_{\mathrm{peak}}/m\omega$, shape functions $f_i(\phi)$ and transverse polarisation vectors $\epsilon_i^\mu$ obeying $\epsilon_i \cdot \epsilon_j = - \delta_{ij}$, so that  $a^{\prime\mu}= a_0 \,  f_i(\phi)\epsilon_i^\mu$. With this, $h$ can be written in the form $h = 1 + \er a_0^2 I_2$, cf.\ (\ref{H}), where $I_2$ is a dimensionless integral of order {\it at most} the pulse duration in $\phi$, i.e.\  $0 \le I_2 \le 2\pi N$. Here, $N$ denotes the number of cycles in the pulse, so we can approximate $I_2 \sim O(N)$. The important parameter is therefore $R := \er a_0^2 N$~\cite{DiPiazza:2009zz}.  One may hence compensate for the smallness of $\alpha$ and $\omega/m$ in (\ref{r}) by using high intensity and/or long pulses~\cite{Foot2}. The regime dominated by purely classical RR without quantum `contamination' is defined by the inequality $\er \ll R \simeq 1$.

\begin{figure}[t!!]
\includegraphics[width=\columnwidth]{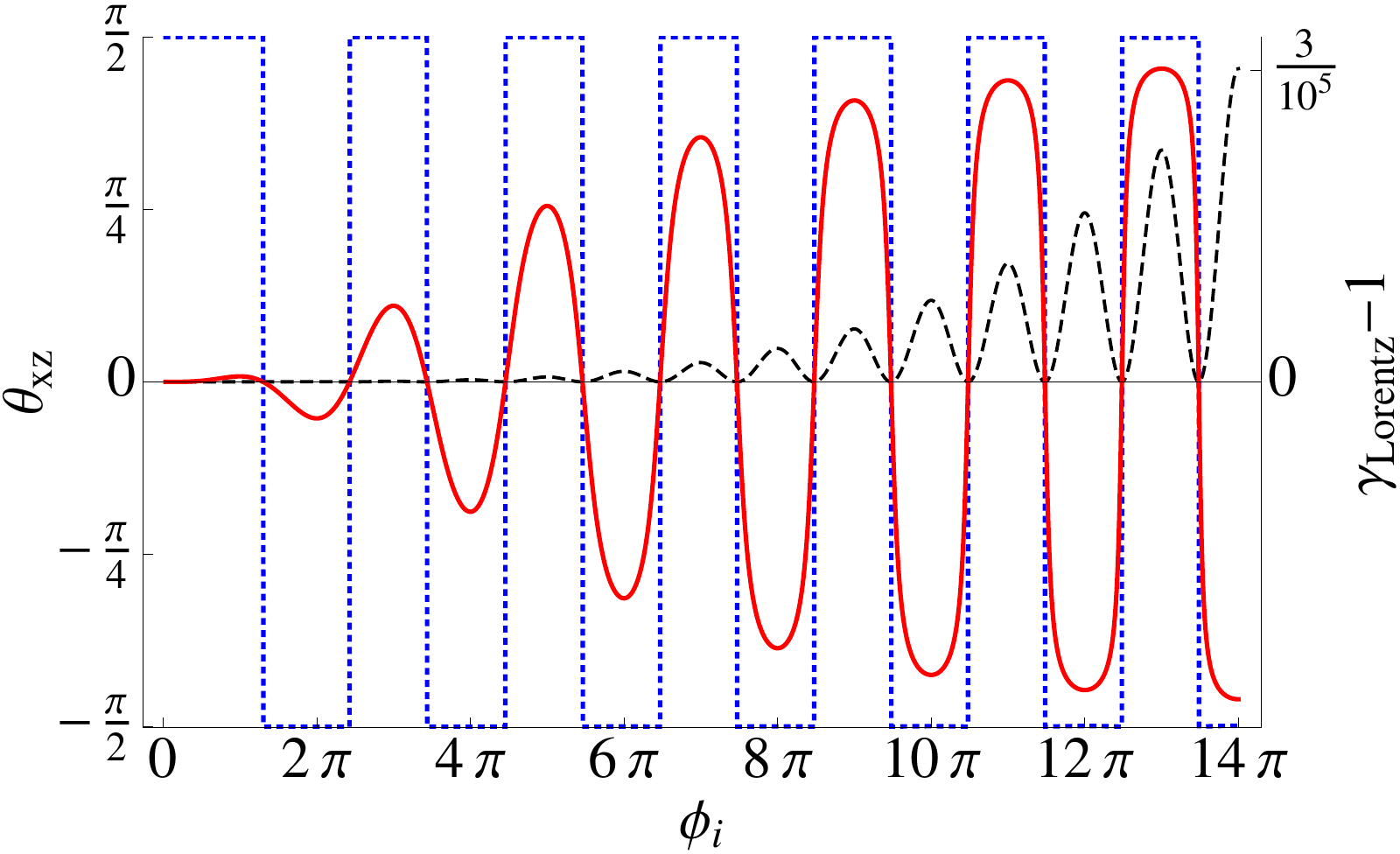}
\caption{\label{FIG:1600} {\it Left scale}: planar angle $\theta_{xz}$ for $N=1600$, $a_0=10$, $\sin^2$ envelope, as a function of ionisation time $\phi_\mathrm{i}$, for Lorentz (blue/dotted) and RR (red/solid). {\it Right}: final Lorentz force gamma for the emitted particles (black/dashed).}
\end{figure}

The size of RR effects increases quadratically with field, and linearly with pulse length. Given that high intensity pulses are formed by tight focussing, it is perhaps best to avoid higher intensities when discussing plane waves with their infinite transverse extent. We will therefore consider, for the most part, long pulses with, by modern standards, moderate intensities. We now present some examples. We chose linear polarisation and a sinusoidal envelope of compact support~\cite{Bardsley:1989zz,Mackenroth:2010jk},
\be \label{MACPULSE}
   \frac{eE^x(\phi)}{m\omega}= a_0 \sin^K (\phi/2N) \sin(\phi) \; , \quad 0 \le \phi 
  \le 2\pi N \; .
\ee
We take the parameters of~\cite{Neitz}: $a_0=10$, $N=1600$ and $K=2$, respectively corresponding to an intensity of $\sim10^{20}$ W/cm$^2$, a total pulse duration of $\sim 4$~ps at optical frequency $\omega\sim 1$~eV, and a $\sin^2$ envelope. The results are shown in Fig.~\ref{FIG:1600}; for linear polarisation, the problem is planar and therefore we plot the angle $\theta_{xz}=\arctan \left(u^x/u^z\right)$ from the positive ($\theta_{xz}=\pi/2$) to negative ($-\pi/2$) $x$-axis. In the Lorentz case, for small $\phii$, the emission direction is almost transverse to the laser, with a small $u^z$ component, so $\theta_{xz}\sim~\pm\pi/2$, with the jumps corresponding to the transverse velocity $u^x$ changing sign while $u^z$ stays small and positive. (This is the reason for plotting $\theta_{xz}$ instead of $\theta$; it allows us to keep track of these sign changes.) For ionisation times within the first few cycles of the pulse, RR can give a change in angle as large as $90^\circ$. The difference between the Lorentz and RR prediction increases with {\it decreasing} ionisation time, so that RR effects are most significant for electrons released in the earliest part of the pulse. Experimentally, one would therefore like a target with a low ionisation threshold. The difference in angle is most significant for those particles which exit the pulse with the least energy; for the $\phii$ in Fig.~\ref{FIG:1600}, the final gamma factor with RR differs from unity by one part in $\mathcal{O}(10^5)$, and differs from the Lorentz force gamma by one part in $\mathcal{O}(10^7)$. (We have only plotted the Lorentz result in Fig.~\ref{FIG:1600}; we return to this shortly.) One would therefore like a clean environment in order that these electrons not be deflected before being detected.

This leads us to a further, striking signature of RR. Note that in Fig.~\ref{FIG:1600}, the transverse Lorentz and RR velocities change sign at the same phases. For sufficiently small ionisation time, though, RR effects can be such that the emission angle $\theta_{xz}\simeq \pi/2$ of the Lorentz case changes to $\theta_{xz}\simeq -\pi/2$ in the RR case; in other words, a particle which would emerge travelling slowly in the positive $x$-direction according to Lorentz, should emerge traveling slowly in the negative $x$-direction according to RR, a $180^\circ$ change in direction.

\begin{figure}[t!!]
\includegraphics[height=0.6\columnwidth]{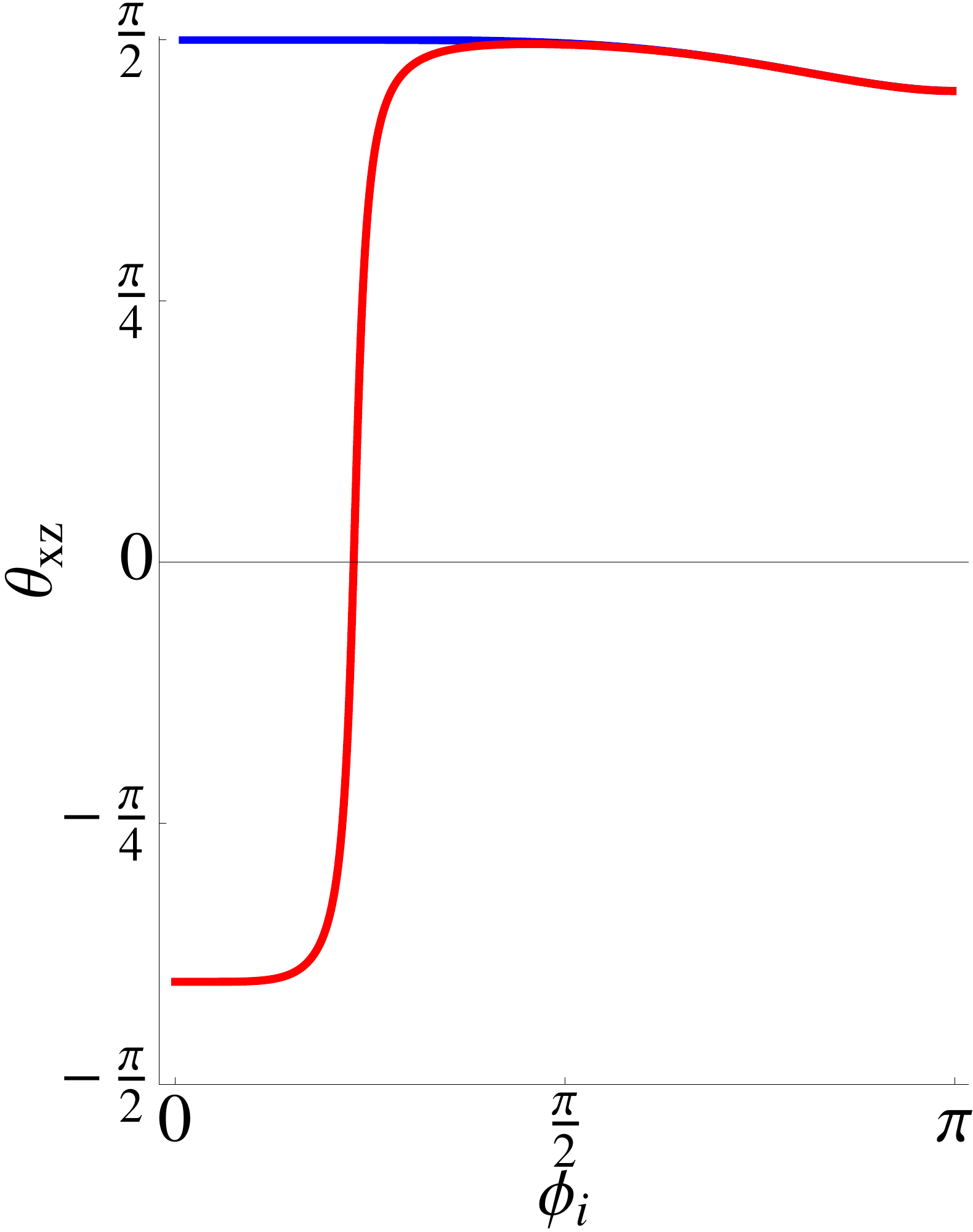}\includegraphics[height=0.6\columnwidth]{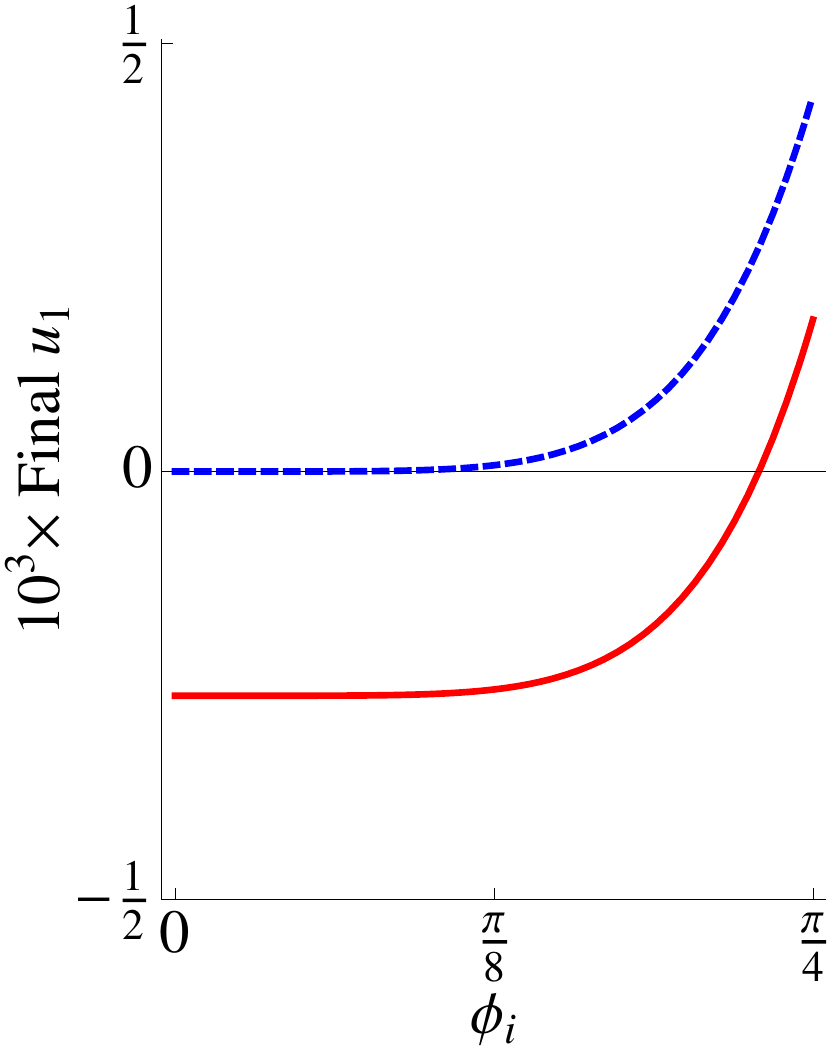}
\caption{\label{FIG:8} {\it Left}: planar angle $\theta_{xz}$, for $N=8$ cycles of an $a_0=50$ pulse with $\sin^4$ envelope. {\it Right}: final transverse velocity, as a function of $\phi_\mathrm{i}$. The sign flip in the transverse component, for $\phi_\mathrm{i}\lesssim {\pi}/{4}$,} is responsible for the large change in emission angle.
\end{figure}

To provide a concrete example, we take a short pulse of $N=8$ cycles, $K=4$ and $a_0=50$. Given the discussion above, this example should not be expected to match a realistic short, focussed pulse, but it is nevertheless interesting to look at the physics involved. The angle $\theta_{xz}$ is plotted in the first panel of Fig.~\ref{FIG:8}. For electrons released early in the first cycle, we see the almost $180^\circ$ change in direction in the emission direction due to RR. This is because, for $\phi_\mathrm{i}\lesssim\pi/4$, RR causes a sign flip in the transverse velocity $u^x$, see the second panel of Fig.~\ref{FIG:8}. Note in particular that the RR contribution to the velocity components is {\it dominant}, with the Lorentz force contribution almost vanishing. The $u^z$ components (not shown) remain positive, with the RR result also dominating. This is an example of the most dramatic deviations from the Lorentz force, which occur for electrons created at special values of the ionisation phase $\phii>0$ such that $a^\mu(\phi_\mathrm{i};\phif)=0$, or almost so. In the absence of RR these electrons gain no net energy and therefore come to rest after leaving the pulse, never reaching the detector to be observed. On the other hand, when RR is taken into account, it will provide the \emph{leading} contribution to the electron final velocity and energy~\cite{Foot3}. Let us illustrate what happens using the polar angle of~(\ref{TAN}). Write the four-velocity $u^\mu$ as a Lorentz term $u_\mathrm{L}^\mu$ and a deviation $\delta u^\mu$ proportional to $R$ (plus, in principle, higher orders), so that $u = u_\mathrm{L} + R \delta u$. The explicit expressions for $u_\mathrm{L}$ and $\delta u$ are easily found from (\ref{LORENTZ.SOLN}) and (\ref{LL.SOLN}), but are not revealing. We can distinguish two cases, the first being `typical', where $a^\mu(\phii,\phif)\not=0$ and then
\be\label{TAN-GENERIC}
	\tan^2\theta_{\mathrm{RR}} = \tan^2\theta_{\mathrm{L}}\bigg(1 + 2R\bigg[\frac{u_\mathrm{L}^x\delta u^x}{u_\mathrm{L}^x u_\mathrm{L}^x}-\frac{\delta u^z}{u_\mathrm{L}^z}\bigg]\bigg) \;,
\ee
so that RR gives a small, $\mathcal{O}(R)$, correction to the Lorentz result as expected.  However, in the case that the Lorentz contribution vanishes, $a^\mu(\phii;\phif)=0$, we have instead
\be\label{TAN-SPECIFIC}
	\tan^2\theta_{\mathrm{RR}} = \frac{\delta u^x\delta u^x}{\delta u^z\delta u^z} \;,
\ee
which, it is important to stress, is independent of $R$ and $\er$. We have thus found observables where the total effects due to the Lorentz force cancel and {\it only} RR effects remain. This is not in contradiction to the assumption that RR effects are small: for the parameters in this paper on has $R\lesssim 10^{-3}$, and it is easily verified that RR contributions to the velocity components are subleading at each \emph{instant} in time (`local' effects). However, the accumulative (`nonlocal') effect due to RR can still dominate over Lorentz force effects, due to cancellations in the latter.

\begin{figure}[t!]
\includegraphics[width=\columnwidth]{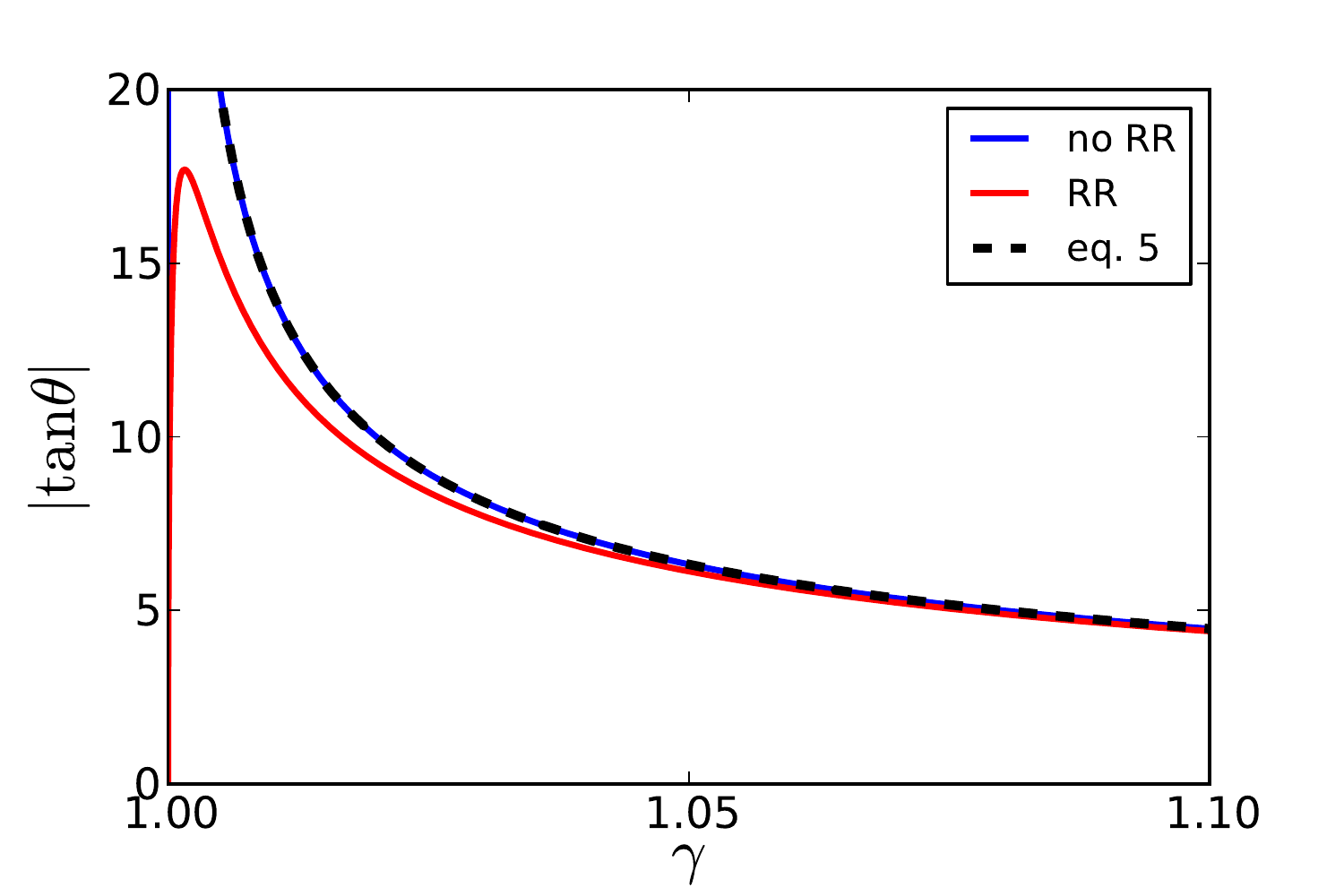}
\caption{\label{FIG:SIM} Numerical result for the correlation between final emission angle and gamma, for $a_0 = 200$, $\sin^4$ envelope. Blue/top curve: Lorentz force, see~\cite{Meyerhofer:1996}. Red: RR result. The black/dashed line corresponds to (\ref{TAN.FTM1}).}
\end{figure}

\paragraph{Discussion:--}
Inspired by the successful experiment \cite{Meyerhofer:1996}, and using the same plane wave model, we have identified a parameter regime in which RR effects are leading rather than subleading. We are aware, though, that numerical methods will be essential for extending the above to more refined models~\cite{Narozhny:1996, Bulanov:2004de,Gonoskov:2012FieldModel,rmp,Pfeiffer}. As a preparation for this, we have performed numerical simulations using the code PATRA \cite{Rykovanov:2013}. For a given charged particle, the code solves the Landau-Lifshitz equation using a fourth order Runge-Kutta method. To mimic ionisation, each particle is assigned a certain unique value of the electric field amplitude, below which the particle is immobile. When the field exceeds this amplitude, the particle is `injected' into the simulations (with zero velocity) and begins to move under the influence of the EM field. The code reproduces the first panel plot of Fig.~\ref{FIG:8} extremely well; the respective curves are on top of each other. In Fig.~\ref{FIG:SIM} we plot, using the code, the parametric relations between final gamma and emission angle. The blue (top) curve is the Lorentz result, as in (\ref{TAN.FTM1}) and \cite{Meyerhofer:1996}. The red curve shows the RR result, with the difference being greatest for smallest final gamma. As for experimental realisation of the proposed scheme, external guiding structures and high-order modes for laser pulses~\cite{BELLA, rmp} can be used to counter diffraction of laser radiation and prevent ponderomotive scattering, ensuring the interaction of electrons with only the high intensity part of the laser pulse. Moreover, the utilisation of a gas with ionisation threshold of the order of the peak pulse intensity, and the employment of a pulse profile such that the ionisation probability is maximal at the phase $\phii$ for which $a^\mu(\phii;\phif)=0$, should enhance the observable effects of RR.
\paragraph{Conclusions:--}
We have described a simple experiment which can be used to observe the effects of classical radiation reaction, without going to ultra-high intensities. As in the earlier experiment~\cite{Meyerhofer:1996}, a target is ionised by a laser pulse, and the final electron momenta are measured. It is not necessary to measure the emitted radiation. The data can be used to distinguish between radiating and non-radiating equations of motion, which predict different values for the final electron momenta. The essential signal is the appearance of low energy electrons scattered at angles forbidden by the Lorentz force equation.

 The sign-flip signal discussed above would of course be a particularly clear signal of RR, but arranging for this to to be visible in a realistic experiment will require fine tuning. The `generic' signal, that the electron emission angle changes, however, is robust. For long pulses at moderate intensity, for which the transverse focussing is not too tight, the plane wave model should give a reasonably accurate first approximation.

\paragraph{{Acknowledgements.}}
The authors are supported by EPSRC, grant EP/I029206/1-YOTTA (C.~H.), the European Research Council, contract 204059-QPQV (A.I.\ and M.M.), the Swedish Research Council contract 2011-4221 (A.I.), the National Science Foundation under grant PHY-0935197, and the Office of Science of the U.S. Department of Energy under contracts DE-AC02-05CH11231 and DE-FG02-12ER41798.


\begin{thebibliography}{99}
\bibitem{Foot1}This is a natural notation since we are looking at the integral of the electric field, but no gauge potential is needed or used.
\bibitem{Foot2} One could also utilise large initial gamma factors, $\gamma_{\mathrm{i}} \gg 1$, but this would take us out of the classical regime. This is not the scenario we are pursuing, so we do not discuss this case further.
\bibitem{Foot3} The same would be true for electrons already present before the arrival of the pulse (i.e., $\phii \le 0$), but this is not the situation we consider here.
\end{thebibliography}
\end{document}